\documentclass[twoside]{dis04}
\def\runauthor{C. Aidala for the PHENIX Collaboration}
\def\shorttitle{$A_{N}$ for Neutral Pions and Charged Hadrons from PHENIX}
\begin{document}

\title{Single-Spin Transverse Asymmetry in Neutral Pion and Charged Hadron Production at PHENIX}

\author{C. Aidala for the PHENIX Collaboration}

\address{Columbia University\\
538 W. 120th Street, New York, NY 10027\\
U.S.A\\
E-mail: caidala@bnl.gov }

\maketitle

\abstracts{The PHENIX experiment at the Relativistic Heavy Ion Collider (RHIC) at 
Brookhaven National Laboratory (BNL) has measured the single-spin transverse asymmetry ($A_{N}$) 
for neutral pion and non-identified charged hadron production at $x_{F}\sim 0$ over a 
transverse momentum range of $\sim$ 0.5 to 5.0~GeV/c from polarized proton-proton interactions 
at a center of mass energy ($\sqrt{s}$) of 200~GeV.  The asymmetries observed are consistent with 
zero.}

\section{Introduction}

Contrary to original expectations from perturbative quantum chromodynamics (pQCD) \cite{Kane}, 
large transverse single-spin asymmetries have been observed in a number of experiments 
\cite{E704,HERMES,STAR}, ranging in energy from 
$\sqrt{s}$~=~20-200~GeV.  The large 
asymmetries seen have stimulated more careful study by the theoretical community of polarized 
cross sections, in particular their dependence on the intrinsic transverse momentum 
of the partons ($k_{T}$) (see e.g. \cite{Mulders}).  

Over the years, a number of models based on pQCD have been developed to predict these $k_{T}$ 
dependencies and to explain the observed asymmetries.  Among these models are the Sivers 
effect \cite{Sivers}, transversity and the Collins effect \cite{Collins}, and various models 
which attribute the observed asymmetries to higher-twist contributions (see e.g. \cite{QS}).  

The unpolarized cross sections for mid-rapidity production as well as forward production of 
neutral pions have been measured in 200-GeV proton-proton collisions at RHIC and have been 
found to agree well with next-to-leading order (NLO) pQCD calculations \cite{PHENIXpi0,STAR}.  
This agreement indicates that NLO pQCD will be applicable in interpreting polarized data from RHIC 
as well and provides a solid theoretical foundation for the spin physics program.  

The analyzing power ($A_{N}$) is the azimuthal asymmetry in particle production by a transversely 
polarized beam on an unpolarized target.  Experimentally, the analyzing power on the left side 
of the beam is given by 
\begin{equation}
A_{N} = \frac{1}{P_{beam}}\frac{1}{<|cos \phi|>}\frac{N^{\uparrow} - RN^{\downarrow}}{N^{\uparrow} + RN^{\downarrow}}
\label{asymmetry1}
\end{equation} 
where $P_{beam}$ is the beam polarization, $<|cos \phi|> = \frac{\sum_{j=1}^{N}|cos \phi_{j}|}{N}$ 
a correction for the azimuthal detector acceptance, $N^{\uparrow}$ ($N^{\downarrow}$) 
the experimental yield from up- (down-) polarized bunches, and 
$R = \frac{L^{\uparrow}}{L^{\downarrow}}$ the relative luminosity of and up- and 
down-polarized bunches. 

\section{Data and Analysis}
The data analyzed were taken during the first polarized proton run at RHIC (2001-2), in which 
approximately 0.15~$pb^{-1}$ were collected at PHENIX.  The stable spin direction of the 
protons is vertical.  Both beams were polarized, and then 
single-spin analyses were performed by averaging over the spin states of one beam.  The average 
beam polarization was $15\%$, measured using proton-carbon 
elastic scattering in the coulomb nuclear interference (CNI) region \cite{CNIPol}.  
The analyzing power was measured at a beam energy of 22~GeV to within $\pm30\%$ \cite{CNIMeas} 
and is here estimated to be the same at 100~GeV.  

In PHENIX, a collision trigger is provided by the coincidence of signals in two beam-beam counters 
(BBCs) \cite{NIMInner}.  Events within 75~cm of the nominal interaction point were taken as 
minimum bias (MB) events.  The BBCs were also used to determine the relative luminosity 
between bunches of opposite polarization sign.  

\begin{figure}[!thb]
\vspace*{6.0cm}
\begin{center}
\includegraphics{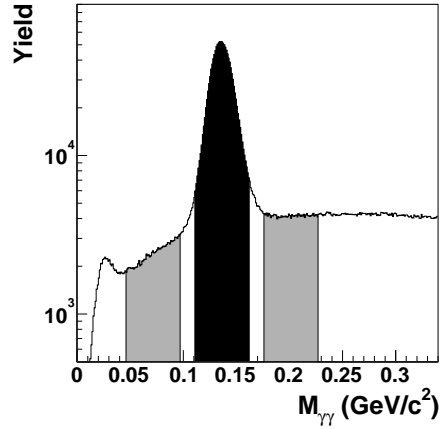}
\caption[*]{Two-photon invariant mass distribution.  The analyzing power was first calculated 
for the region including both signal and background under the $\pi^{0}$ mass peak, shown in black.  
Then the contribution of the combinatorial background to the asymmetry under the peak was estimated 
and corrected for using the grey sidebands.}
\label{fig:mass}
\end{center}
\end{figure}

Neutral pions were reconstructed via their decay to two photons using finely granulated 
electromagnetic calorimeters \cite{NIMEMCal}.  A high-energy photon trigger with an energy threshold of 
approximately 0.8~GeV, in coincidence with the MB trigger, was used to collect the 
neutral pion data \cite{PHENIXpi0}.   
Neutral pion yields were extracted by integrating the two-photon invariant mass spectrum 
from 0.12-0.16~$GeV/c^{2}$, as indicated by the black band in Figure~\ref{fig:mass}.  
 The contribution from combinatorial background ranged from 
58$\%$ to 9$\%$ in the lowest and highest transverse momentum bins.  The contribution to 
the asymmetry by the combinatorial background under the peak was estimated by 
calculating the asymmetry of the grey bands on both sides of the 
signal (Fig. \ref{fig:mass}).  The asymmetry under the peak was then corrected using 

\begin{equation}
A_{N}^{\pi^{0}} = \frac{A_{N} - rA_{N}^{BG}}{1-r} \quad
\sigma_{A_{N}^{\pi^{0}}} = \frac{\sqrt{\sigma^{2}_{A_{N}} + r^{2}\sigma^{2}_{A_{N}^{BG}}}}{1-r}
\label{correctedAsym}
\end{equation}
where $r$ is the fraction of combinatorial background under the peak.

Tracks of charged hadrons were reconstructed using a drift chamber and one of several pad chambers 
\cite{NIMTracking} as well as the BBCs to determine the collision vertex.  In order to eliminate 
electrons from photon conversions, it was required that there be 
no hit in the ring-imaging Cherenkov detector (RICH) \cite{NIMPID}.  
The electron contamination in the final data sample was less than 1$\%$.  The decay background from 
long-lived particles could not be eliminated.  However, because these tracks have incorrectly 
reconstructed momentum, the analyzing power for these tracks is $p_{T}$ independent.  

The asymmetry was determined for each fill using Eq.~\ref{asymmetry1}, then fit to a constant 
across all fills.  The $\chi^{2}$ of the fit and a ``bunch-shuffling'' technique were used 
to check the uncertainties assigned to the asymmetry.  In the bunch shuffling technique, the 
spin direction of each bunch is randomly reassigned and $A_{N}$ is subsequently 
recalculated.  This procedure was repeated many times (1000), and 
the widths of the resulting asymmetry distributions were no larger than the statistical 
uncertainties assigned to the physical asymmetries, indicating that all non-correlated 
bunch-to-bunch and fill-to-fill systematic uncertainties were much smaller than the statistical 
uncertainties.  

\begin{figure}[!thb]
\vspace*{6.0cm}
\begin{center}
\includegraphics{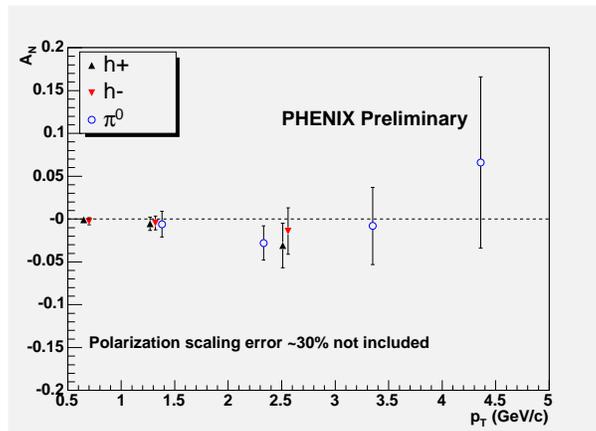}
\caption[*]{Transverse single-spin asymmetries for inclusive charged hadrons and neutral pions.}
\label{fig:asymPlot}
\end{center}
\end{figure}

The resulting asymmetries are shown in Fig.~\ref{fig:asymPlot}.  Systematic uncertainties are 
negligible compared to the statistical uncertainties given.  A total scale uncertainty of 
approximately $\pm30\%$ from the measurement of the beam polarization is not shown.  

\section{Conclusions}

The transverse single-spin asymmetries observed for production of both neutral pions and inclusive 
charged hadrons at $x_{F}\sim0$ are consistent with zero over the measured transverse momentum 
range.  A small asymmetry in this kinematic region follows the trend of previous results, 
which indicate a decreasing asymmetry at decreasing $x_{F}$ \cite{E704Central,STAR}.  
As a significant fraction of particle production in this 
kinematic region comes from gluon scattering, any contribution to the asymmetry from transversity 
and the Collins effect would be suppressed, while contributions from the Sivers effect or other 
mechanisms would remain a possibility.  Further theoretical study of the results will have to be 
performed in order to interpret their full implications for the transverse spin structure of the 
proton.

\section*{Acknowledgements} 
PHENIX acknowledges support from the Department of Energy and NSF (U.S.A.), MEXT and JSPS (Japan), 
CNPq and FAPESP (Brazil), NSFC (China), CNRS-IN2P3 and CEA (France), BMBF, DAAD, and AvH (Germany), 
OTKA (Hungary), DAE and DST (India), ISF (Israel), KRF and CHEP (Korea), RAS, RMAE, and RMS (Russia), 
VR and KAW (Sweden), U.S.CRDF for the FSU, US-Hungarian NSF-OTKA-MTA, and US-Israel BSF.

\end{document}